\documentclass[review]{elsarticle}

\usepackage{graphicx} 
\usepackage{lineno}
\usepackage{hyperref}
\usepackage{comment} 
\usepackage{xcolor}

\modulolinenumbers[1]
\newcommand{\LambdaHO}{$\Lambda_{pO}$}
\newcommand{\muH}{$\mu p$}
\newcommand{\muO}{$\mu O$}
\newcommand{\hfs}{$\Delta E_{hfs}$}
\newcommand{\LaBr}{$LaBr_3(Ce)$}
\newcommand{\Ka}{$K_{\alpha}$}
\newcommand{\Kb}{$K_{\beta}$}
\newcommand{\Kg}{$K_{\gamma}$}
\newcommand{\xray}{\mbox{X-ray}}
\newcommand{\xrays}{\mbox{X-rays}}
\newcommand{\dtcut}{\mbox{$\Delta t_s > 30$ ns}}

\begin{document}

\begin{frontmatter}

\title{Measurement of the muon transfer rate from muonic hydrogen to oxygen in the range 70-336~K}

\author[INFN-Ts]{C.~Pizzolotto}
\author[INFN-Ts,Ing-Ud]{A.~Sbrizzi \corref{mycorrespondingauthor}}
\cortext[mycorrespondingauthor]{Corresponding author}
\ead{Antonio.Sbrizzi@cern.ch}
\author[Poland]{A.~Adamczak}
\author[Bulgaria]{D.~Bakalov} 
\author[INFN-Bo,Uni-Bo]{G.~Baldazzi }
\author[INFN-Ts,Uni-Ud]{M.~Baruzzo }
\author[INFN-Bicocca,Scienze-Mi]{R.~Benocci }
\author[INFN-Bicocca]{R.~Bertoni }
\author[INFN-Bicocca,Fisica-Mi]{M.~Bonesini} 
\author[INFN-Ts,Abdus]{H.~Cabrera} 
\author[INFN-Ts,Uni-Ud]{D.~Cirrincione}
\author[INFN-Bicocca,Fisica-Mi]{M.~Clemenza} 
\author[INFN-Rm,Ing-Rm]{L.~Colace}
\author[INFN-Ts,Elettra]{M.~Danailov} 
\author[Bulgaria]{P.~Danev} 
\author[Uni-Pv,INFN-Pv]{A.~de~Bari}
\author[INFN-Pv]{C.~De~Vecchi }
\author[INFN-Rm,Uni-Rm]{M.~De~Vincenzi} 
\author[Uni-Na]{E.~Fasci}
\author[INFN-Bo,INAF-Bo]{F.~Fuschino}
\author[INFN-Ts,Abdus,Togo]{K.~S.~Gadedjisso-Tossou}
\author[Uni-Na]{L.~Gianfrani}
\author[Riken]{K.~Ishida}
\author[INFN-Bo,INAF-Bo]{C.~Labanti}
\author[INFN-Bicocca,Scienze-Mi]{V.~Maggi}
\author[INFN-Bicocca]{R.~Mazza}
\author[Uni-Pv,INFN-Pv]{A.~Menegolli}
\author[INFN-Ts]{E.~Mocchiutti}
\author[INFN-Ts,Uni-Ud]{S.~Monzani}
\author[Uni-Na]{L.~Moretti}
\author[INFN-Bo,INAF-Bo]{G.~Morgante}
\author[Abdus]{J.~Niemela}
\author[INFN-Mi,Fisica-Mi-Celoria]{A.~Pullia}
\author[INFN-Mi,cnr-Mi]{R.~Ramponi}
\author[INFN-Bo]{L.~P.~Rignanese}
\author[INFN-Pv]{M.~Rossella}
\author[Bulgaria]{M.~Stoilov}
\author[INFN-Ts]{L.~Stoychev\fnref{myfootnote}}
\fntext[myfootnote]{Current address: Institute of Solid State Physics, Bulgarian Academy of Sciences, 72, Tzarigradsko Chaussee, Blvd., 1784 Sofia, Bulgaria}
\author[INFN-Ts]{J.~J.~Su\'arez-Vargas}
\author[INFN-Rm]{L.~Tortora}
\author[INFN-Bicocca]{E.~Vallazza}
\author[INFN-Ts,Uni-Ud,Riken]{A.~Vacchi}

\address[INFN-Ts]{Sezione INFN di Trieste, via A. Valerio 2, Trieste, Italy}%formerNumbering%1
\address[Ing-Ud]{Dipartimento Politecnico di Ingegneria e Architettura dell'Universit\`a di Udine, via delle Scienze 206, Udine, Italy}
\address[Poland]{Institute of Nuclear Physics, Polish Academy of Sciences, Radzikowskiego 152, PL31342 Krak\'{o}w, Poland}%2
\address[Bulgaria]{Institute for Nuclear Research and Nuclear Energy,
Bulgarian Academy of Sciences, blvd.\ Tsarigradsko ch.~72, Sofia 1784, Bulgaria}%3
\address[INFN-Bo]{Sezione INFN di Bologna, viale Berti Pichat 6/2, Bologna, Italy}%4
\address[Uni-Bo]{Dipartimento di Fisica ed Astronomia, Universit\`a di Bologna, via Irnerio 46, Bologna, Italy}
\address[Uni-Ud]{Dipartimento di Scienze Matematiche, Informatiche e
   Fisiche, Universit\`a di Udine, via delle Scienze 206, Udine, Italy}%5
\address[INFN-Bicocca]{Sezione INFN di Milano Bicocca, Piazza della Scienza 3, Milano, Italy}%6
\address[Scienze-Mi]{Dipartimento di Scienze dell'Ambiente e della Terra,   Universit\`a di Milano Bicocca, Piazza della Scienza 1, Milano,   Italy}%7
\address[Fisica-Mi]{Dipartimento di Fisica G. Occhialini, Universit\`a di   Milano Bicocca, Piazza della Scienza 3, Milano, Italy}%8
\address[INFN-Rm]{Sezione INFN di Roma Tre, Via della Vasca   Navale 84, Roma, Italy}%9
\address[Ing-Rm]{Dipartimento di Ingegneria, Universit\`a degli Studi
   Roma Tre, Via V. Volterra 62, Roma, Italy}%10
\address[Elettra]{   Sincrotrone Elettra Trieste, SS14, km 163.5, Basovizza, Italy}%11
\address[Uni-Pv]{Dipartimento di Fisica, Universit\`a di Pavia, via
   A.~Bassi 6, Pavia, Italy}%12
\address[INFN-Pv]{Sezione INFN di Pavia, Via A.~Bassi 6, Pavia,
   Italy}%13
\address[Uni-Rm]{Dipartimento di Matematica e Fisica, Universit\`a di Roma Tre, Via della Vasca Navale 84, Roma, Italy}%14
\address[Uni-Na]{Sezione INFN di Napoli e Dipartimento di Matematica e Fisica, Universit\`a della Campania “Luigi Vanvitelli”, Viale Lincoln 5, Caserta, Italy}%15
\address[INAF-Bo]{INAF-OAS Bologna, %Area della Ricerca,
   via P.~Gobetti 93/3, Bologna, Italy}%16
\address[Abdus]{The Abdus Salam International Centre for
   Theoretical Physics, Strada Costiera 11, Trieste, Italy}%17
\address[Togo]{Laboratoire de Physique des Composants \`a
   Semi-conducteurs (LPCS), D\'epartment de physique, Universit\'e de
   Lom\'e, Lom\'e, Togo}%18
%\address[GranSasso]{Gran Sasso Science Institute, via F. Crispi 7, L'Aquila, Italy}%19
%\address[ISIS]{ISIS Neutron and Muon Source, STFC Rutherford-Appleton Laboratory, Didcot, OX11 0QX, United Kingdom}%20
\address[Riken]{Riken Nishina Center, RIKEN, 2-1   Hirosawa, Wako, Saitama 351-0198, Japan}%21
%\address[INO]{INO-CNR, via Madonna del Piano 10, 50019 Sesto Fiorentino, Italy}%22
%\address[IFAC-CNR]{IFAC-CNR, via Madonna del Piano 10, 50019 Sesto Fiorentino, Italy}%23
\address[INFN-Mi]{Sezione INFN di Milano, via Celoria 16, Milano, Italy}%24
\address[Fisica-Mi-Celoria]{Dipartimento di Fisica, Universit\`a degli Studi di Milano, via Celoria 16, Milano, Italy}%25
\address[cnr-Mi]{IFN-CNR, Dipartimento di Fisica, Politecnico di   Milano, piazza Leonardo da Vinci 32, Milano, Italy}%26
%\address[India]{Indian Centre for Space Physics, Kolkata, India}%27
%\address[Dalian]{Dalian Institute of Chemical Physics of the Chinese Academy of Sciences, 457 Zhongshan Road, Dalian 116023, P. R. China}%28

\begin{abstract}

The first measurement of the temperature dependence of the muon transfer rate from muonic hydrogen to oxygen was performed by the FAMU collaboration in 2016.
The results provide evidence that the transfer rate rises with the temperature in the range 104-300~K.
This paper presents the results of the experiment done in 2018 to extend the measurements towards lower (70~K) and higher (336~K) temperatures. 
The 2018 results confirm the temperature dependence of \LambdaHO{} observed in 2016 and sets firm ground for comparison with the theoretical predictions.

\end{abstract}


\end{frontmatter}

%\linenumbers

\section{Introduction}

The goal of the FAMU experiment is to extract the Zemach radius of the proton, with an accuracy better than 1\%, from a measurement of the hyperfine splitting of muonic hydrogen ground state (\hfs)\cite{Mocchiutti_2018}.

The experiment consists in counting the number of muon transfers from muonic hydrogen (\muH) to oxygen (\muO) when a low energy muon beam 
stops in a hydrogen target containing a fraction of oxygen of the order of 1\% (by weight). 
The target is contained in a high reflectivity optical cavity where an intense laser with finely tunable frequency is injected.
After the formation of \muH{} atoms, the muon transfer process leads to the creation of excited \muO{} atoms whose de-excitation cascade gives rise to the \Ka, \Kb{} and \Kg{} spectral lines of muonic oxygen (133~keV, 158~keV and 167~keV) which provide the signature of the muon transfer process. 
The muon transfer probability is larger when muonic hydrogen has a higher thermal energy, as confirmed by our recent measurement \cite{Emiliano}.
If the laser is tuned to the right hyperfine splitting transition energy, the \muH{} atoms, predominantly occupying the lower singlet spin state, will be excited to the triplet state.
When de-excited in collision with the surrounding $H_2$ molecules to the singlet state, muonic hydrogen acquires kinetic energy due to the non-radiative de-excitation process, which translates in a larger muon transfer probability.
By tuning the laser wavelength on the maximum number of detected \xrays, it is possible to provide a precise measurement of \hfs.

The knowledge of the muon transfer rate from \muH{} to \muO{} (\LambdaHO) is important to optimize the experimental conditions for the measurement of \hfs\cite{Cecilia}.

Since 2013, four preliminary measurements without the laser system have been performed in preparation for the spectroscopic data taking.
In 2016, the collaboration dedicated an entire data taking session to the measurement of \LambdaHO{} as a function of temperature in the range 104-300~K\cite{Emiliano}. 
The same experimental setup was used in March and December 2018 to extend the measurements down to 70~K and up to 336~K.
This paper presents the results of the analysis performed to extract the temperature dependence of \LambdaHO{} from 2018 data.

\section{Experimental setup}

The FAMU experiment is performed at the RIKEN-RAL\cite{RAL} facility which provides a pulsed-muon beam with a repetition rate of 50~Hz.
Each bunch consists of two gaussian muon spills (FWHM~=~70~ns) separated by about 320~ns.
In order to maximize the probability of muonic hydrogen formation in the target, the muon beam momentum was set to 55~MeV/c in the 2018 data taking. The average muon rate at 55~MeV/c is about 3$\times10^4$/s.

The RIKEN-RAL facility has four muon beam delivery ports.
The experiment was installed at Port4 in 2016, while Port1 hosted the experiment in 2018.
Compared to Port4, Port1 is better isolated from the external environment and the detectors operate in more stable temperature conditions.

A detailed description of the experimental setup can be found in Ref.\cite{Adam2018}.
The cryogenic target contains an aluminium cylindrical vessel filled with the mixture of hydrogen and oxygen and it is surrounded by different types of \xray{} detectors.
The vessel is internally coated with a thin layer of heavy materials (gold and nickel) to stop outgoing muons. 
The fast nuclear capture of muons in the coating material suppresses the photon background associated to the slow muon capture in the aluminium walls.

The analysis presented in this paper is based on data recorded with six scintillating counters.
Each counter consists of a \LaBr{} cylindrical crystal (1~inch diameter and 1~inch long) coupled to an Hamamatsu \mbox{R11265U-200} photomultiplier. 
Waveforms are sampled with a 14-bit 500 Ms/s CAEN V1730C digitizer and recorded in time windows of 8.19~$\mu$s (12-bit TDC).

The rise time of photomultiplier signals ($\tau_r$) is close to the time bin of the digitizer (2~ns).
In order to improve signal reconstruction, $\tau_r$ is increased with a pulse shaper within the limit of negligible pile-up effects ($\tau_r$~=~16~ns).
In 2016, $\tau_r$ was set to 12~ns. 

The trigger provided by the beam facility was adjusted to start data acquisition about 300~ns before the arrival of the first muon spill.
\section{Data sample} \label{sec:data_sample}

The temperature dependence of \LambdaHO{} is measured by changing the temperature of the cryogenic target system hosting the $H_2/O_2$ mixture.
The gas pressure increases with the temperature because the target is sealed by a valve which keeps a constant gas density inside the target\cite{Emiliano}.
The transfer rate from $\mu p$ atoms was measured at each temperature.
The analysis presented in this paper is based on data recorded during two runs taken in the same experimental conditions, in March and December 2018. 

March 2018 data were taken at four temperatures (272~K, 300~K, 323~K and 336~K). The maximum temperature reachable by the cryogenic target sets the upper limit on the temperature. The cryogenic-cooler helium compressor works up to a pressure of 22.8~bar which corresponds to a temperature of 350~K on the cold head. 
For safety reasons, the temperature of the cold head was kept below 340~K, the target temperature being slightly lower (336~K) due to heat losses along the copper braids connections between the cold head and the target.

\begin{figure}[h]
\includegraphics[scale=0.3]{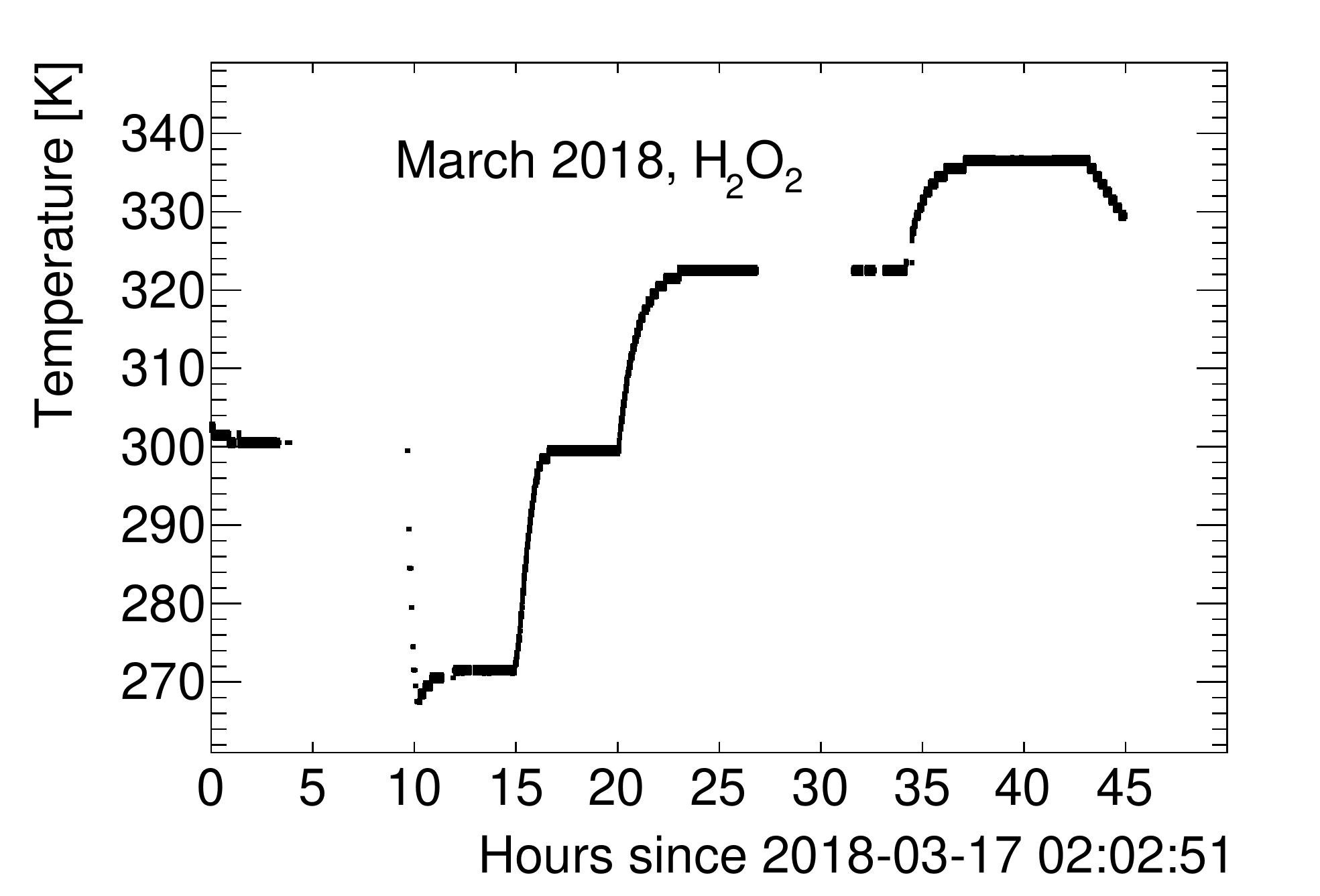}
\includegraphics[scale=0.3]{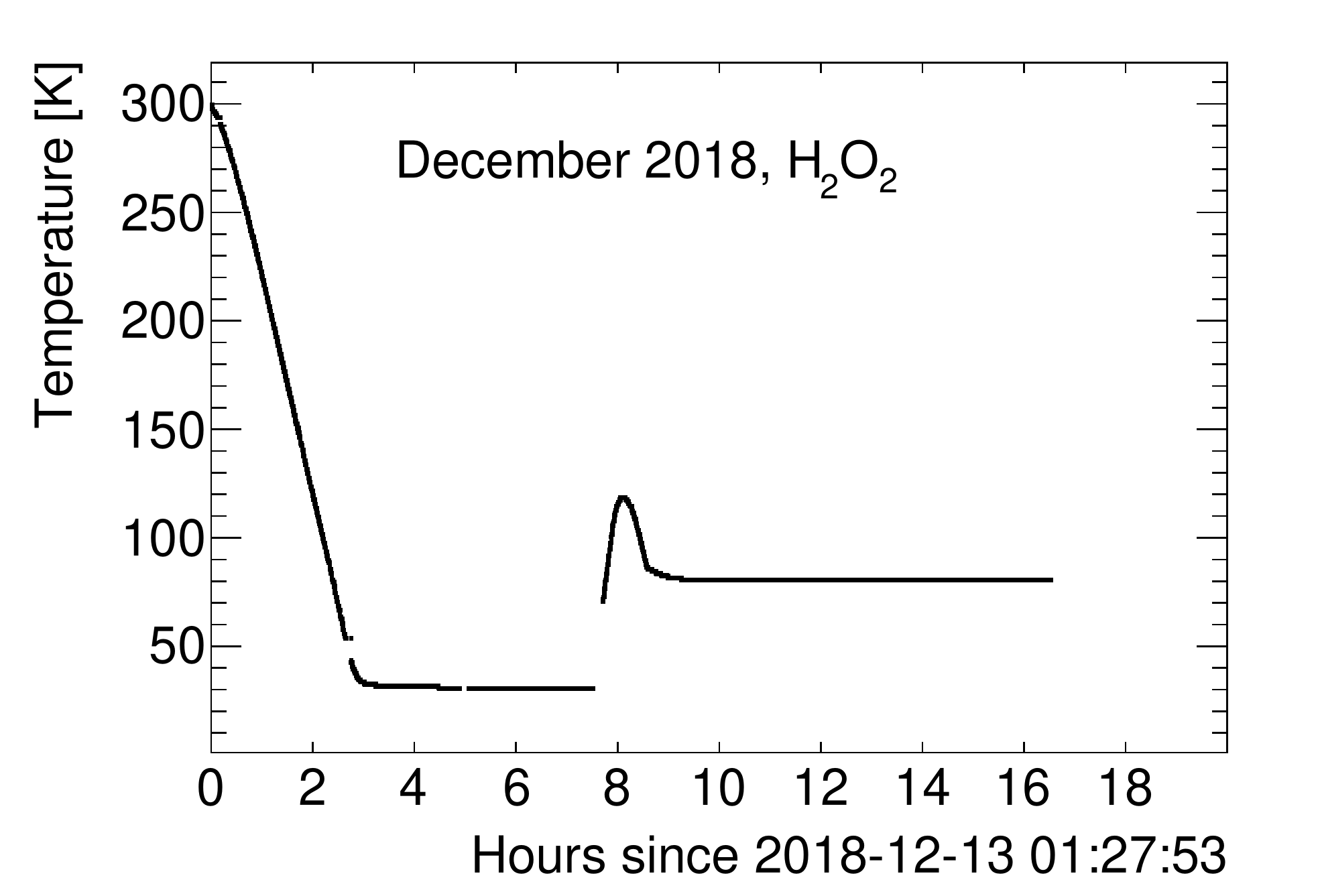}
\centering
\caption{Time-dependence of the $H_2/O_2$ target temperature in March (left) and December (right) 2018 data taking.}
\label{fig:plottempVsTime}
\end{figure}

The cooling system of the final spectroscopy experiment will be based instead on liquid nitrogen in order to minimize the vibrations transmitted to the laser system.
For this reason, December 2018 was mostly devoted to take data at about the liquid nitrogen temperature (80~K).
Data at lower temperatures were taken to explore the region of oxygen condensation.
Figure \ref{fig:plottempVsTime} shows the time-dependence of the $H_2/O_2$ target temperature in the 2018 data taking.  

During the first two hours of data taking in December 2018, the target temperature decreased almost linearly from 300~K to 30~K.

Data taken at variable temperature between 100 and 300 K are used to compare the muon transfer rate measured in 2018 to the one measured in 2016 at constant temperature.
Even though the two measurements are not performed in the same temperature conditions, the comparison is necessary to extract the oxygen concentration in the 2018 target, as it will be explained in Sec.\ref{analysis}.

The March 2018 sample consists of 4.1~M triggers on the $H_2/O_2$ target and 1.1~M on pure hydrogen for background estimation, while the December 2018 sample has 1.3~M triggers on the $H_2/O_2$ target and 0.7~M on pure hydrogen.

\begin{comment}
Fig.~\ref{fig:plottemp_ndet03_000K} shows the total number of triggers as a function of the temperature in 2018. 

\begin{figure}[h]
\includegraphics[scale=0.3]{figures/plottemp_ndet03_000K_mar}
\includegraphics[scale=0.3]{figures/plottemp_ndet03_000K_dec}
\centering
\caption{Total number of muon triggers as a function of the temperature in the runs of March (left) and December (right) 2018.}
\label{fig:plottemp_ndet03_000K}
\end{figure}
\end{comment}

\section{Data analysis}\label{analysis}

\begin{figure}[h]
\includegraphics[scale=0.3]{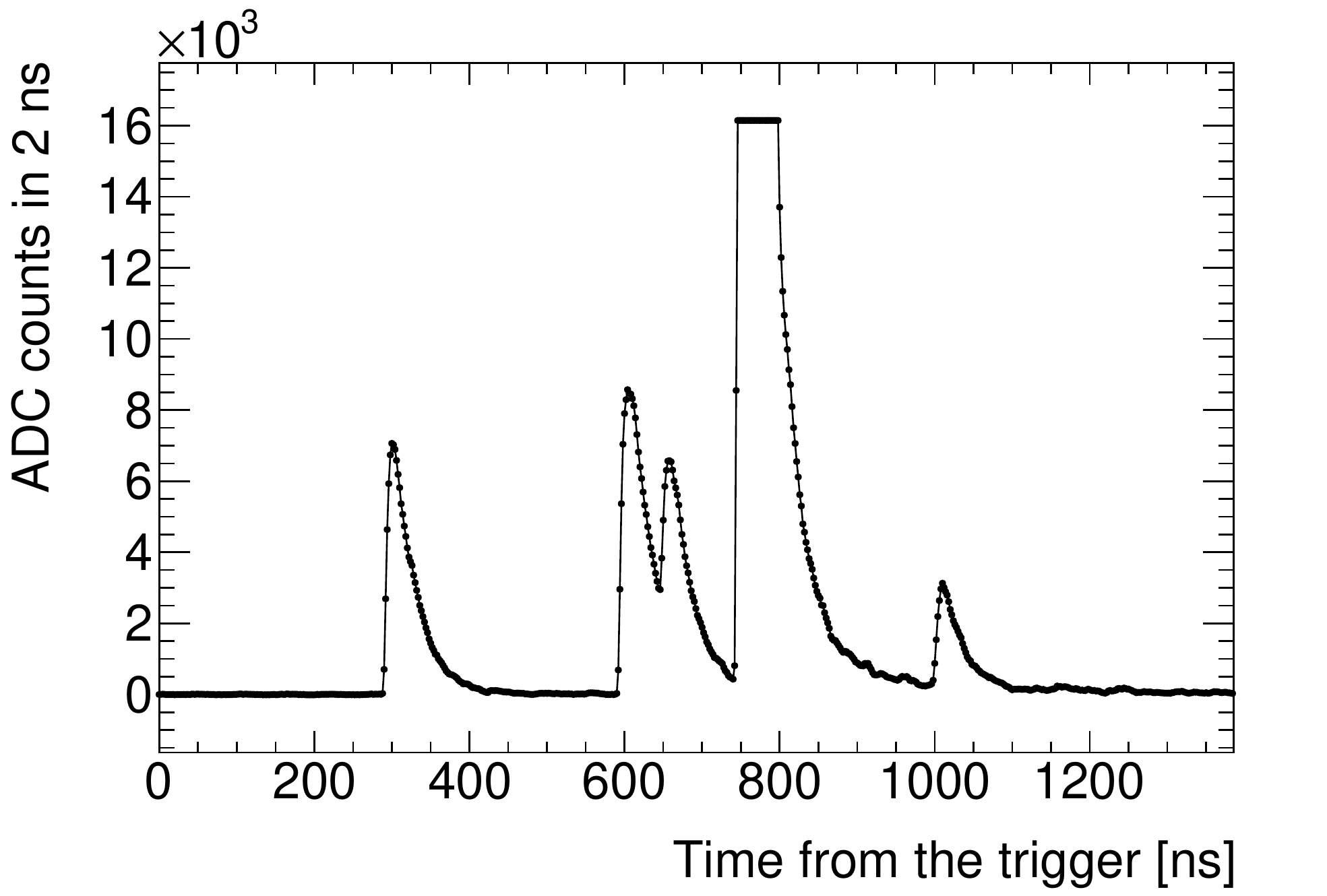}
\includegraphics[scale=0.3]{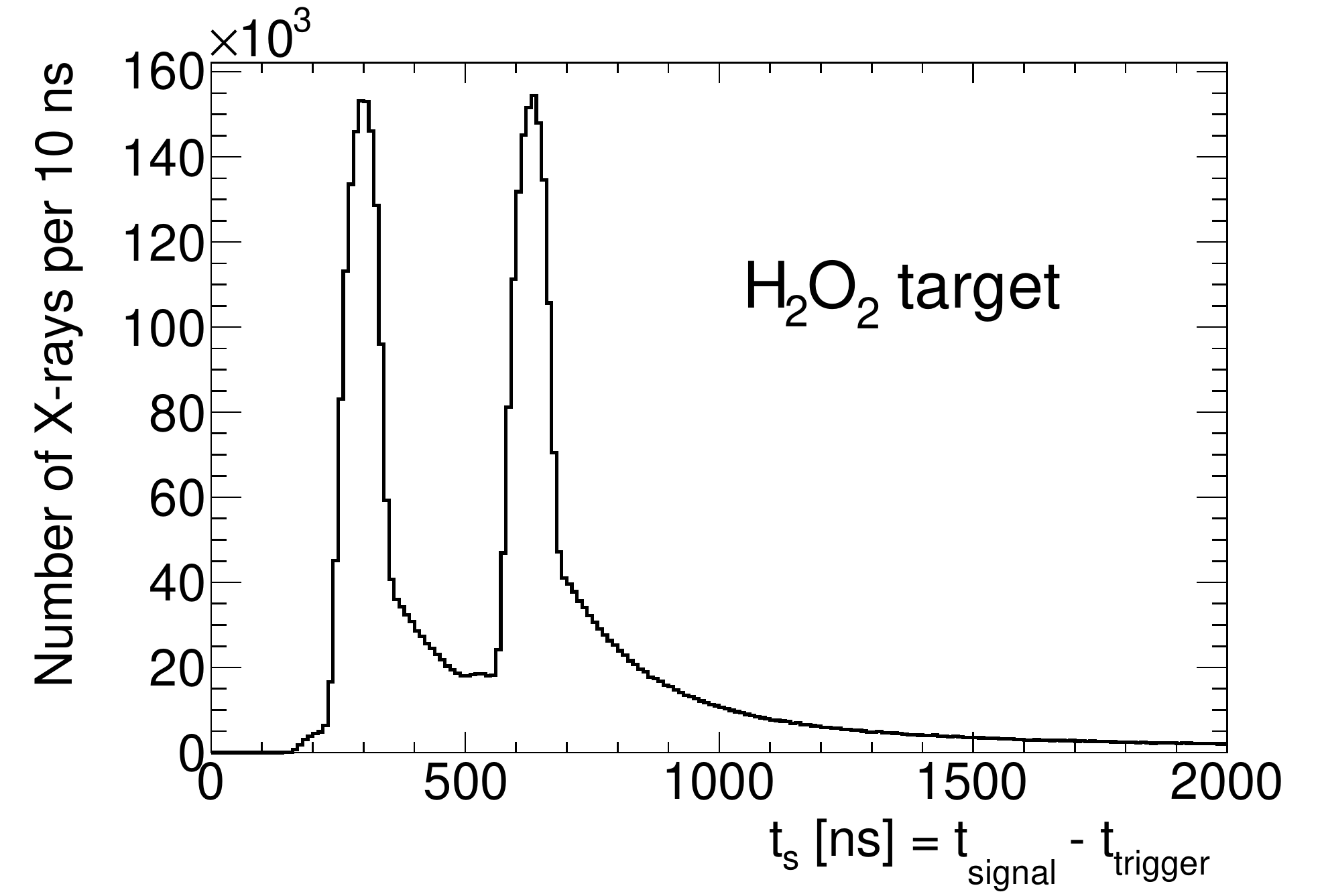}
\centering
\caption{Example of a digitized waveform (left).
Each pulse corresponds to an \xray.
The $1^{st}$ and the $5^{th}$ pulse are well separated, the $2^{nd}$ and the $3^{rd}$ overlap, the $4^{th}$ pulse saturates the FADC counter.
Example of starting time distribution of signals recorded with a \LaBr{} detector when muons are captured in the $H_2/O_2$ target (right).}
\label{fig:cDelta}
\end{figure}

An example of digitized waveform recorded with the FAMU data acquisition system after baseline subtraction is shown in Fig.~\ref{fig:cDelta}.

The starting time of a pulse ($t_s$) is obtained by requiring that the first derivative of the waveform is larger than three times the local average fluctuation above the mean.
The baseline RMS is calculated in the proximity of $t_s$.
Pulses with a baseline RMS larger than 20~ADC counts are rejected to improve the energy resolution.
The right panel in Fig.~\ref{fig:cDelta} shows an example of $t_s$ distribution. 
The two peaks at 300~ns and 630~ns are the prompt \xray{} produced at the arrival of the two muons spills.
Delayed \xrays{} are produced at time larger then the arrival of the second muon spill ($t_s >$ 900~ns).

The pulse amplitude is evaluated at the time in which the first derivative of the waveform goes back to zero.
If the derivative does not cross the zero, the pulse is tagged as unresolved.
Unresolved pulses and pulses lying above unresolved pulses are rejected.
In order to suppress pile-up effects that might spoil the energy resolution of the detector, a minimal time separation between pulses is required (\dtcut).
In case of overlapping pulses, the exponential tail of the previous pulse is subtracted.
Pulses that saturate the FADC counter ($2^{14}$) are rejected.

The pulse amplitude is calibrated in energy by using prompt \xrays{} from elements present in the target, i.e. aluminium (65.8~keV, 88.8~keV, 346~keV), nickel (107~keV, 309~keV). 
Prompt \xrays{} are selected in time windows defined around the arrival time of the two muon spills (240-340~ns and 570-670~ns). 
Delayed \xray{} signals from oxygen (133~keV) and photons from electron-positron annihilation (511~keV) are also used for the energy calibration.

\begin{figure}[h]
\includegraphics[scale=0.3]{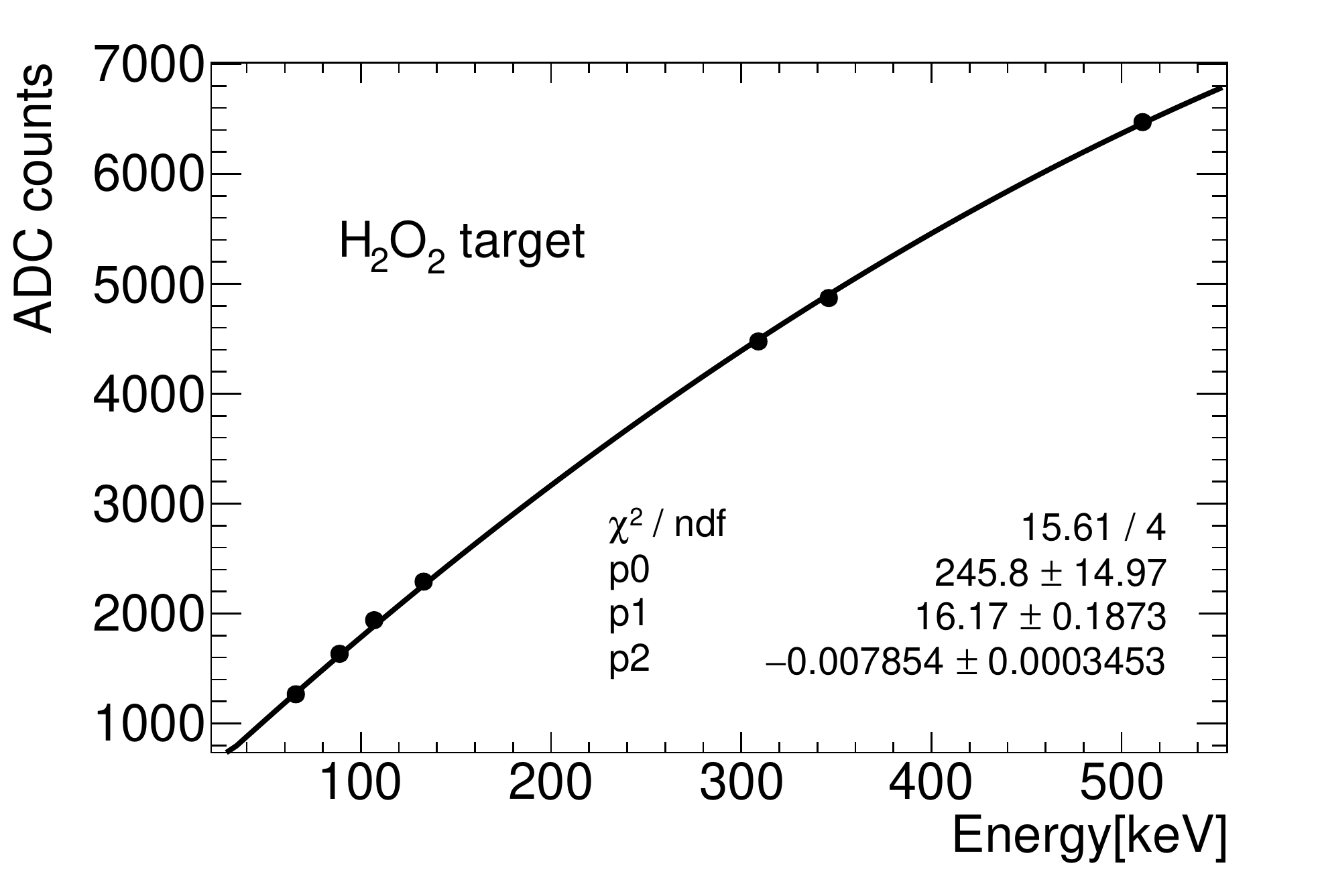}
\centering
\caption{
Calibration curve of a \LaBr{} detector.
Vertical error bars are the squared sum of gaussian mean errors and systematic errors on background modelling and pulse selection.
The fit to a second degree polynomial function is superimposed. Errors are smaller than the marker size.}
\label{fig:plotcal}
\end{figure}

Each calibration point is obtained by fitting the pulse-height spectrum with a combination of a gaussian peak and a functional model for the background in the region of a given emission line.
The systematic errors are evaluated by changing the background description model and the pulse selection criteria.
Figure \ref{fig:plotcal} shows the calibration curve of a \LaBr{} detector used in this analysis.

Figure \ref{fig:plotcal} shows that the relation between amplitude and energy fits well to a second degree polynomial equation.
The energy resolution of the \LaBr{} detector in Fig.~\ref{fig:plotcal} is 10$\%$ FWHM at the \Ka{} line of oxygen (133~keV).

\subsection{Detector live time and selection efficiency}

Detector live time and selection efficiency are calculated with data-driven methods.
The duration of saturated pulses is used to estimate the live time.
The selection efficiency is the fraction of identified and non-saturated pulses that are resolved and far from other pulses (\dtcut).

\begin{figure}[h]
\includegraphics[scale=0.3]{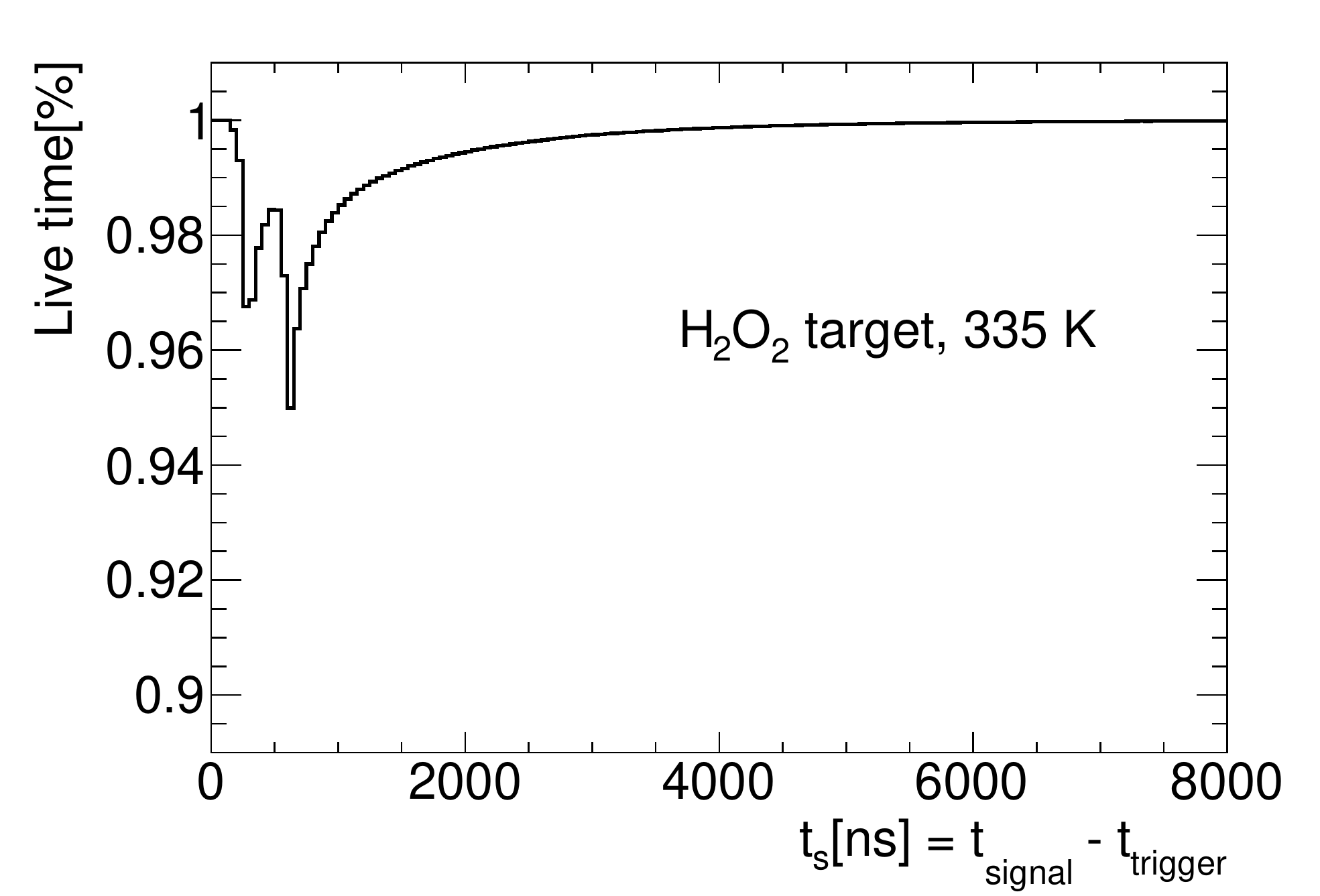}
\includegraphics[scale=0.3]{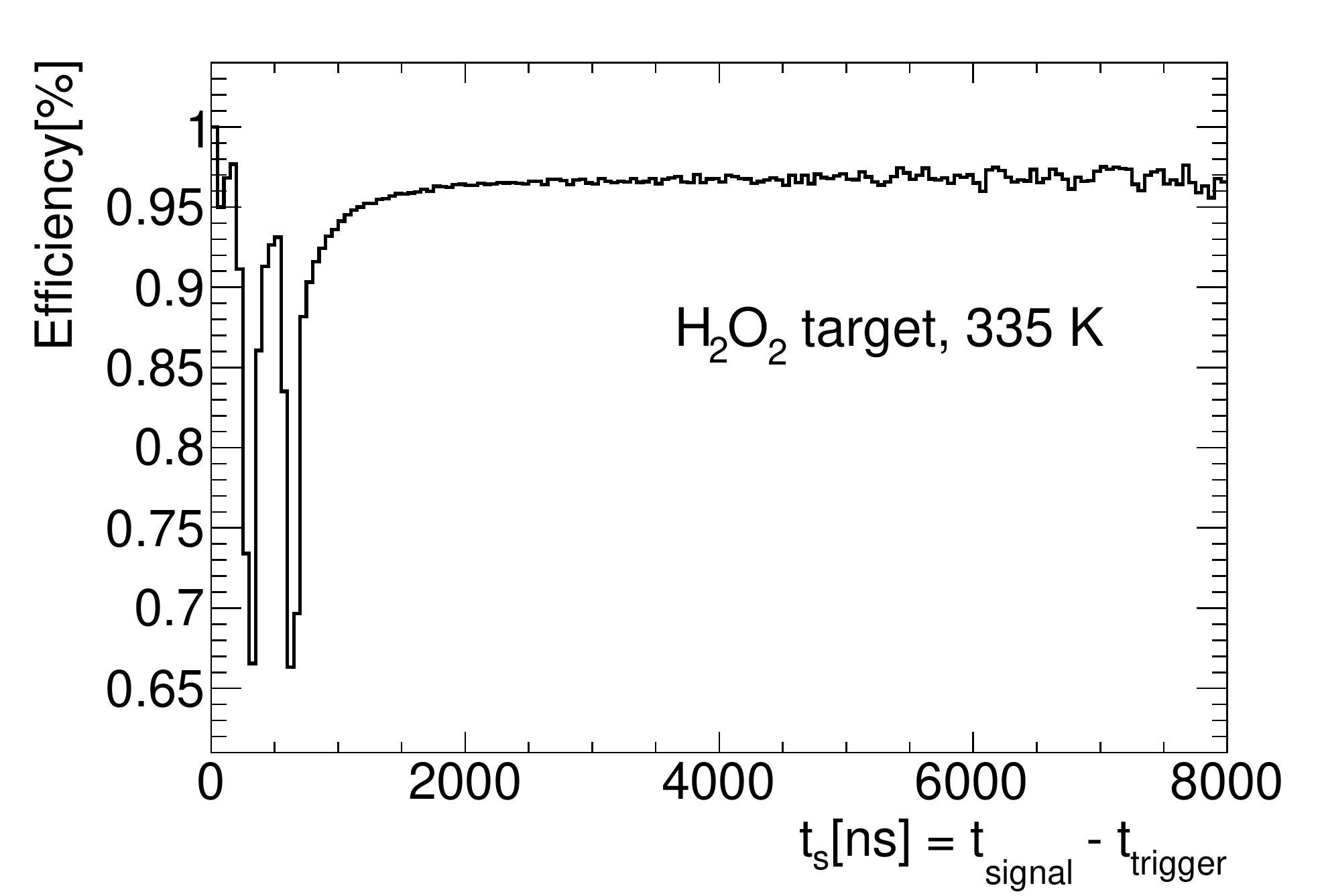}
\centering
\caption{Average live time (left) and selection efficiency (right) at 336~K of the six available \LaBr{} detectors as a function of the time after trigger in 50~ns time bins.}
\label{fig:plotlive_tbinw050ns_335K_ndet10}
\end{figure}

The simulation shows that 99.9$\%$ of the pulses are correctly identified by the reconstruction software and that the cut on $\Delta t_{s}$ suppresses the systematic effects on pulse amplitude determination.
Figure \ref{fig:plotlive_tbinw050ns_335K_ndet10} shows the average live time and the selection efficiency of the six available \LaBr{} detectors as a function of the time after trigger.

Live time and selection efficiency are smaller in the proximity of the two muon spills due to a larger pile-up probability.
Target gas composition has a negligible effect on live time and selection efficiency.

\subsection{\xray{} counting}

The number of muon transfers to oxygen is measured by counting the number of delayed oxygen \xrays{} recorded by \LaBr{} detectors.

Background is evaluated by measuring the number of delayed \xrays{} produced in the $H_2$ target.
According to the simulation, the main source of background is emission of bremsstrahlung photons by electrons originating from muon decays, while the contribution of prompt \xrays{} produced in the target material (aluminium, nickel and gold) is not relevant.

\begin{figure}[ht]
\includegraphics[scale=0.3]{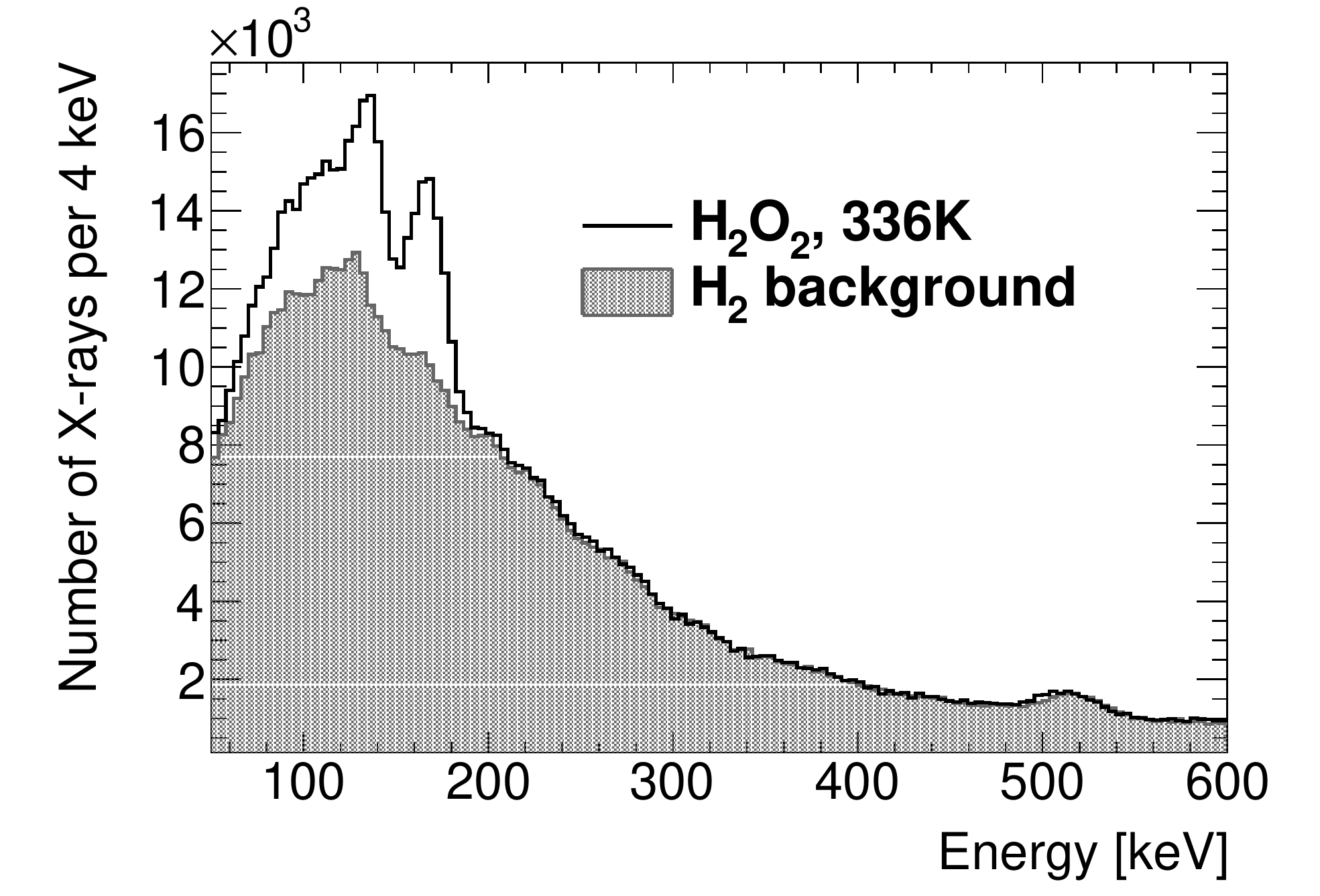}
\includegraphics[scale=0.3]{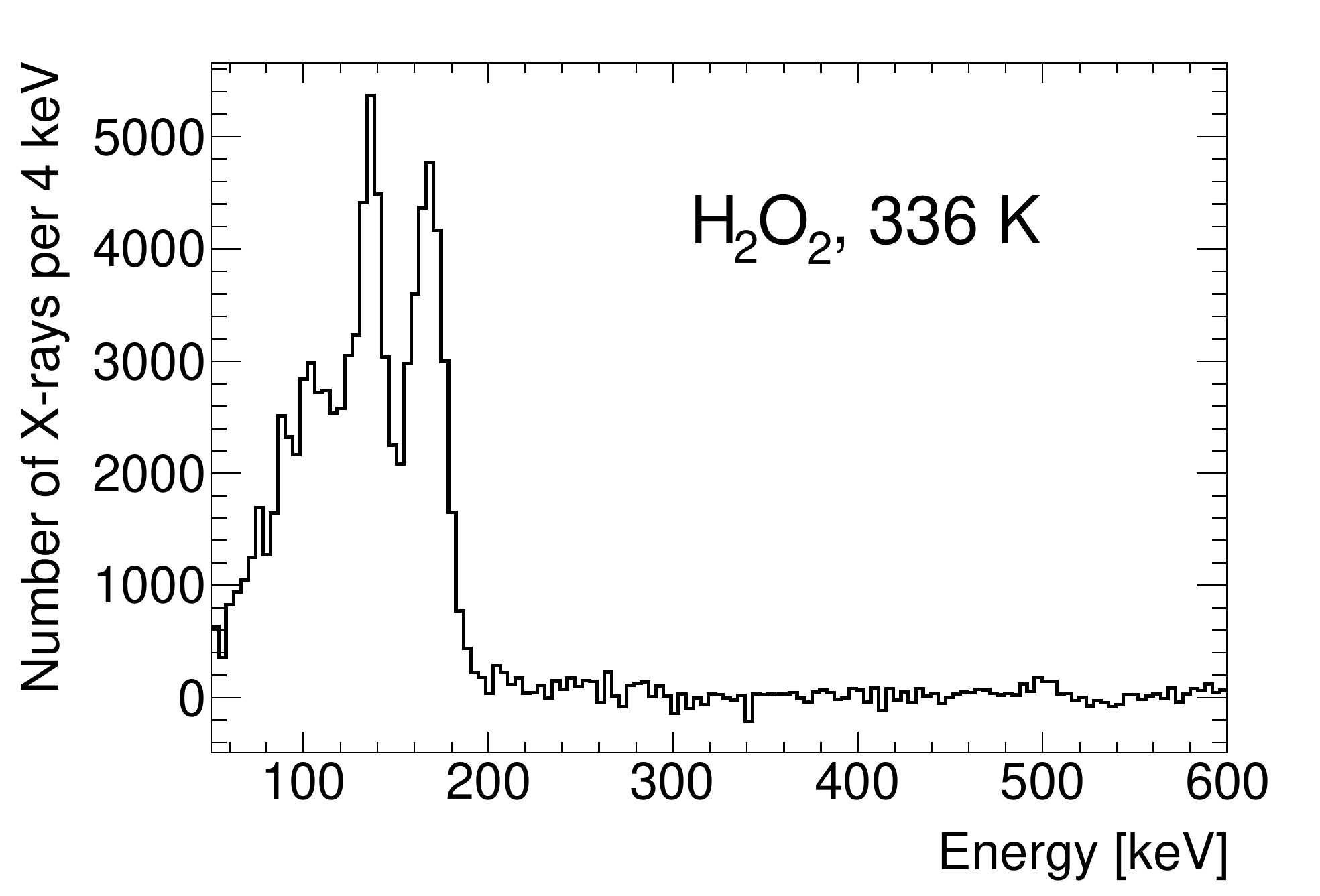}
\centering
\caption{Energy spectra of delayed \xrays{} produced in $H_2/O_2$ and $H_2$ with $t_s$ in the range 900-1200~ns (left). 
Spectra are normalized in the range 250-350~keV. 
The temperature of the $H_2/O_2$ target was 336~K, while pure hydrogen was at 300~K.
Background subtracted energy spectrum (right).}
\label{fig:plotrate_ndet10_336K_bsys0_0900ns_1200ns}
\end{figure}

Figure \ref{fig:plotrate_ndet10_336K_bsys0_0900ns_1200ns} shows the energy spectrum of delayed \xrays{} recorded with all the available \LaBr{} detectors, before and after background subtraction. The normalization of the background sample is done in an energy range without \xray{} lines (250-350~keV).

The signal spectrum shows the \Ka{} line of oxygen and a second peak corresponding to the unresolved \Kb{} and \Kg{} lines.
The signal region to the left of the \Ka{} oxygen line is populated by \xrays{} that deposit only a fraction of their energy in the scintillation counter.

\begin{figure}[h]
\includegraphics[scale=0.3]{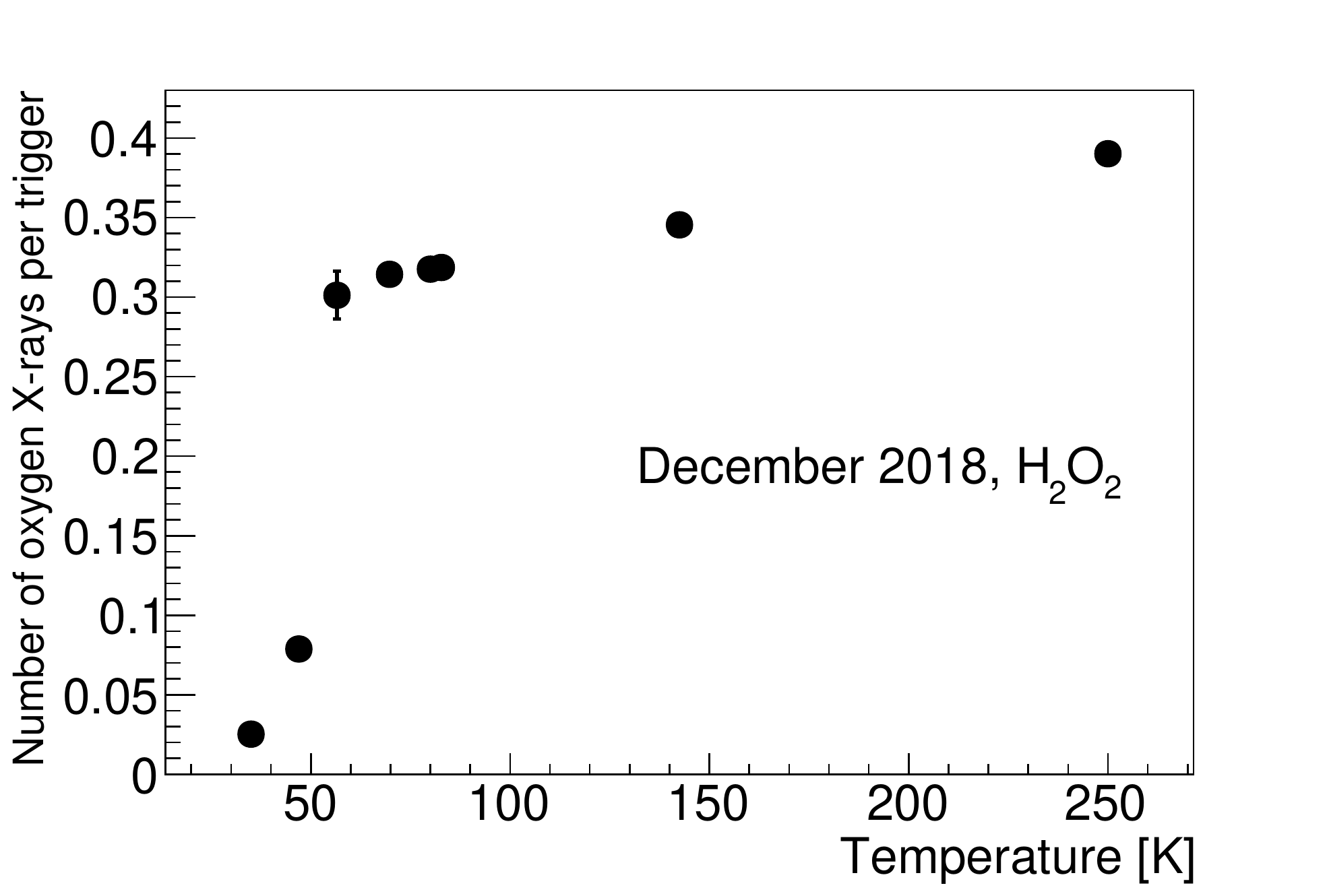}
\centering
\caption{Number of oxygen \xrays{} per trigger in December 2018 as a function of the target temperature.
The temperature of the hydrogen target used to estimate background is 80~K.}
\label{fig:plot_simone}
\end{figure}

Figure \ref{fig:plot_simone} shows the number of oxygen \xrays{} per trigger in December 2018 as a function of the target temperature.

Below the temperature of oxygen condensation (60~K) it is not possible to measure the muon transfer rate to oxygen because the signal goes to zero.

\subsection{Transfer rate measurement} \label{sec:transfer-rate-measurement}

\LambdaHO{} is extracted from the time dependence of the measured number of muon transfers to oxygen in the $H_2/O_2$ target after the thermalization of \muH{} atoms.

At a given temperature T, the number of \muH{} atoms in the target ($N_{\mu p}$) changes with the time $t$ according to the formula:
\begin{equation}
dN_{\mu p}(t) = -N_{\mu p}(t)\lambda_{dis}(T)dt
\end{equation}
The total disappearance rate of \muH{} atoms, $\lambda_{dis}(T)$, can be expressed as:
\begin{equation}
\lambda_{dis}(T) = \lambda_{0} + \phi[c_{p}\Lambda_{pp\mu} + c_{d}\Lambda_{pd}(T) + c_{O}\Lambda_{pO}(T)]
\label{eq:lambda}
\end{equation}
where $\lambda_{0}$ is the disappearance rate of the muons bound to protons, $\Lambda_{pp\mu}$ is the $pp\mu$ formation rate in \muH{} collisions with hydrogen nuclei, $\Lambda_{pd}$ is the muon transfer rate from \muH{} to deuterium, $\Lambda_{pO}$ is the muon transfer rate from \muH{} to oxygen atoms, $\phi$ is the atom density in the gaseous target, $c_{p}$, $c_{d}$ and $c_{O}$ are the hydrogen, deuterium and oxygen atomic concentrations in the target.
The value of the parameters in Eq.~\ref{eq:lambda} and the fitting procedure to extract \LambdaHO{} from the time dependent muon transfers to oxygen are reported in Ref.\cite{Mocchiutti_2018}.
The only difference is the value of $c_{O}$.
In March and December 2018, the target was prepared starting from a gas mixture with an initial oxygen weight concentration\footnote{The oxygen concentration $c_{O}$ is used in units of atomic concentration in the formula of Eq.~\ref{eq:lambda}, and in oxygen weight concentration in the rest of the paper.} of \mbox{$c_{O} = 4.6\%$}. %3028 ppm}
The gas mixture was diluted with hydrogen to reach the ideal conditions in which \mbox{$c_{O} = 0.3\%$} but the procedure was such that the final oxygen concentration in the target exposed to the beam could not be precisely assessed.
Therefore, $c_{O}$ is extracted from data with the procedure reported in Sec.~\ref{sec:cO}.

Assuming a given $c_{O}$ value, \LambdaHO{} is extracted from data by counting the number of \xrays{} in adjacent time bins, starting from 300~ns after the second spill.
For each time bin, the integral of the background subtracted energy spectrum in the range 60-190~keV is corrected by live time and selection efficiency.

\begin{figure}[ht]
\includegraphics[scale=0.3]{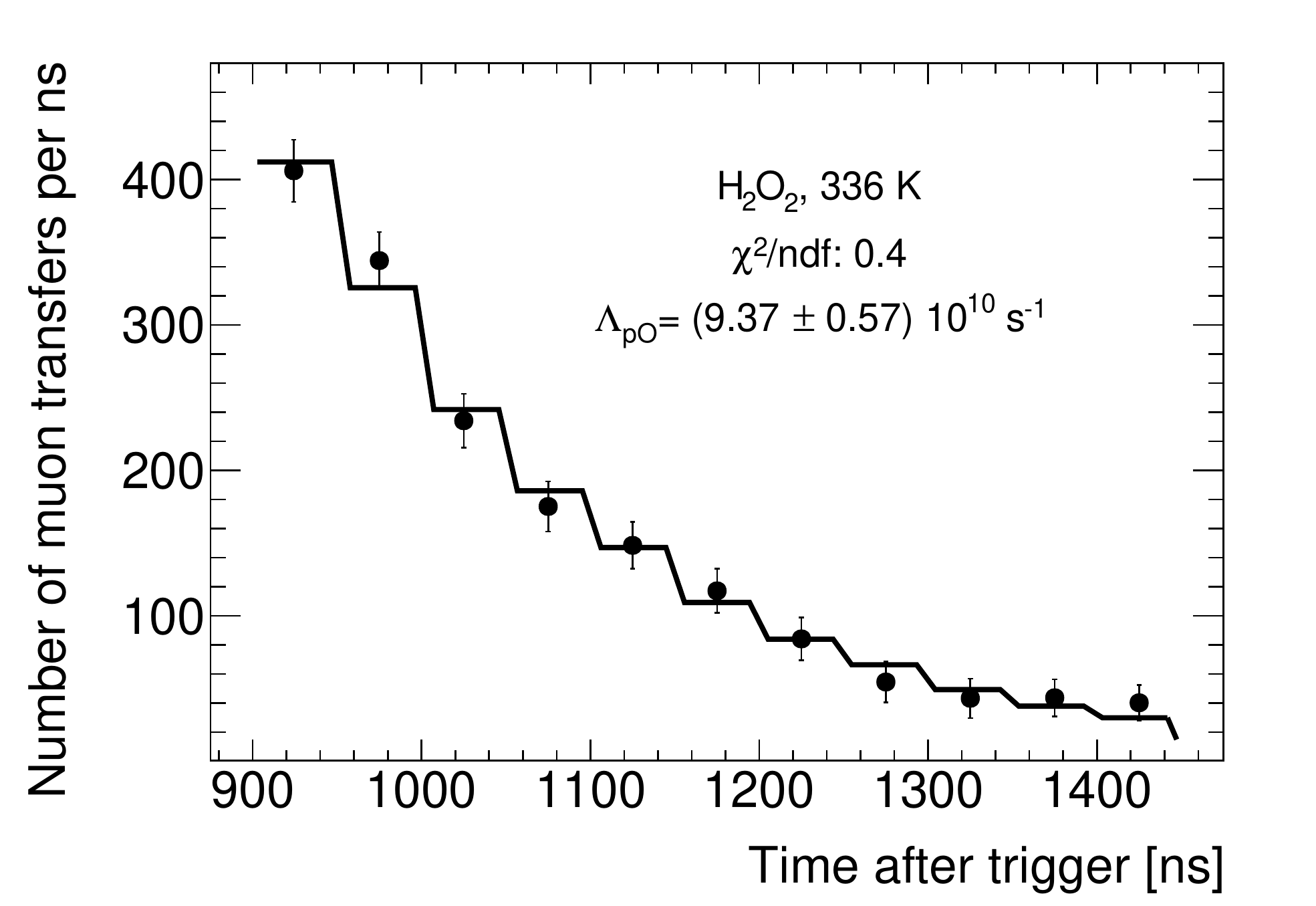}
\includegraphics[scale=0.3]{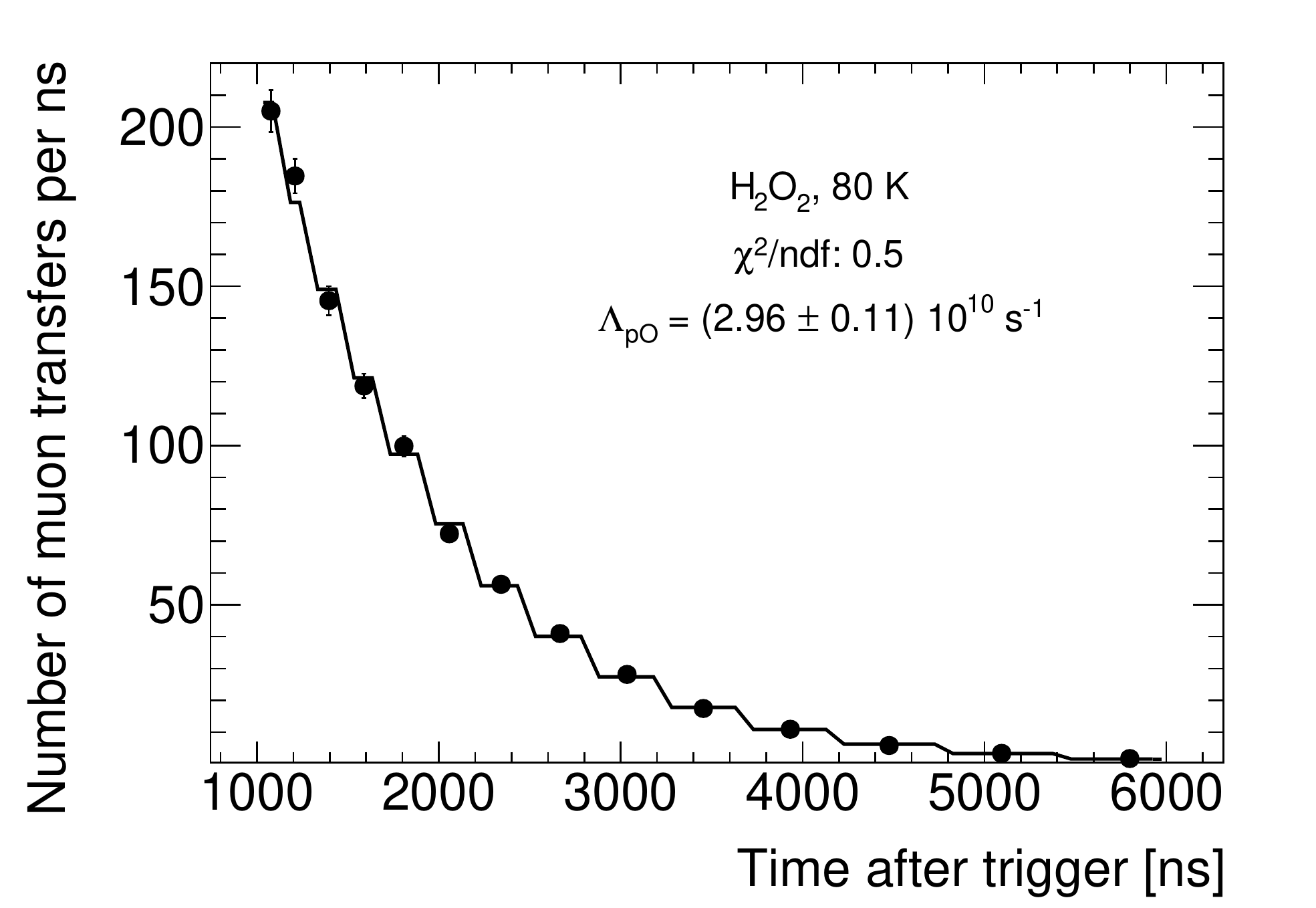}
\centering
\caption{Time dependence of the rate of muon transfers to oxygen measured with all the available \LaBr{} detectors in March-2018 at 336~K (left) and December-2018 at 80~K (right).
The error bars are the sum in quadrature of the statistical and the background-related systematic uncertainties.
The fit to extract \LambdaHO{} is superimposed.}
\label{fig:plotrate_tbinw050ns_336K_ndet10}
\end{figure}

Figure \ref{fig:plotrate_tbinw050ns_336K_ndet10} shows the rate of muon transfers to oxygen measured with all the available \LaBr{} detectors as a function of the time after trigger at 336~K and 80~K.
March 2018 data are divided in 50~ns bins starting from 900~ns. 
December 2018 data are logarithmically binned from 1000~ns on. 

The measurement of \LambdaHO{} is stable against variations of the fitting range towards larger values, which is a proof that prompt \xrays{} background above 900~ns is negligible.

The statistical uncertainty is calculated assuming a Poissonian distribution of the number of muon transfers.
One of the leading sources of systematic uncertainties is background normalization which is evaluated as the maximum variation of the measured number of muon transfers when the subtracted energy spectra undergo a fluctuation of 1$\sigma$ in opposite directions.
Section \ref{sec:cO} is dedicated to the evaluation of the systematic effect on the target gas composition.
Other systematic effects are evaluated in Ref.\cite{Mocchiutti_2018} but they are neglected because their overall contribution is smaller than 1$\%$.

\subsection{Data-driven estimate of the oxygen weight concentration} 
\label{sec:cO}

The value of $c_{O}$ used in Fig.~\ref{fig:plotrate_tbinw050ns_336K_ndet10} is extracted from data by normalising 2016 and 2018 data taken at the same temperatures.

The first step is fitting the transfer rate measured as a function of temperature in 2016\cite{Emiliano} to the lowest order polynomial that well describes the data.
The final choice is a $2^{nd}$ degree polynomial with coefficients $k_{2016}$ (constant term), $k_{1}$ and $k_{2}$ (higher order terms).

The second step is to assume an initial $c_{O}$ value allowing for a first estimate of $\Lambda_{pO}$ at the normalization temperatures: 272~K and 300~K in March, 104~K, 153~K, 201~K, 240~K, and 272~K in December.

The third step is fitting separately March and December 2018 data to the same $2^{nd}$ degree polynomial equation, by letting the constant term $k_{2018}(c_{O})$ as the only degree of freedom ($k_{1}$ and $k_{2}$ are fixed).
The final value of $c_{O}$ results from the minimization of the $\chi^2$ defined as: 
\begin{equation}
\chi^2(c_O) = \frac{[k_{2018}(c_O) - k_{2016}]^2} {\sigma_{2018}^2+\sigma_{2016}^2}
\label{eq:chi2}
\end{equation}
where the $\sigma$ is the error on the $k$ parameter.

\begin{figure}[ht]
\includegraphics[scale=0.3]{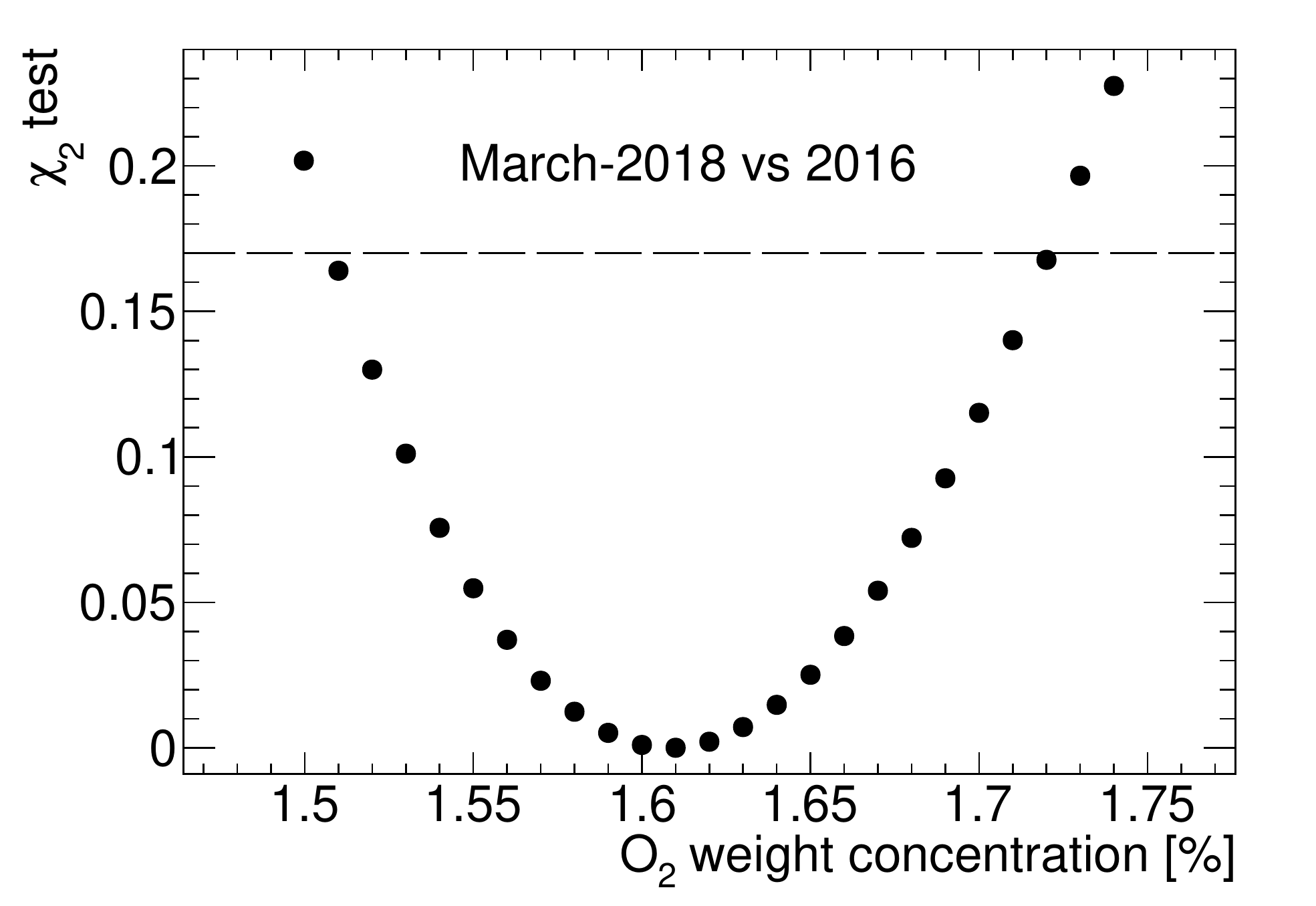}
\includegraphics[scale=0.3]{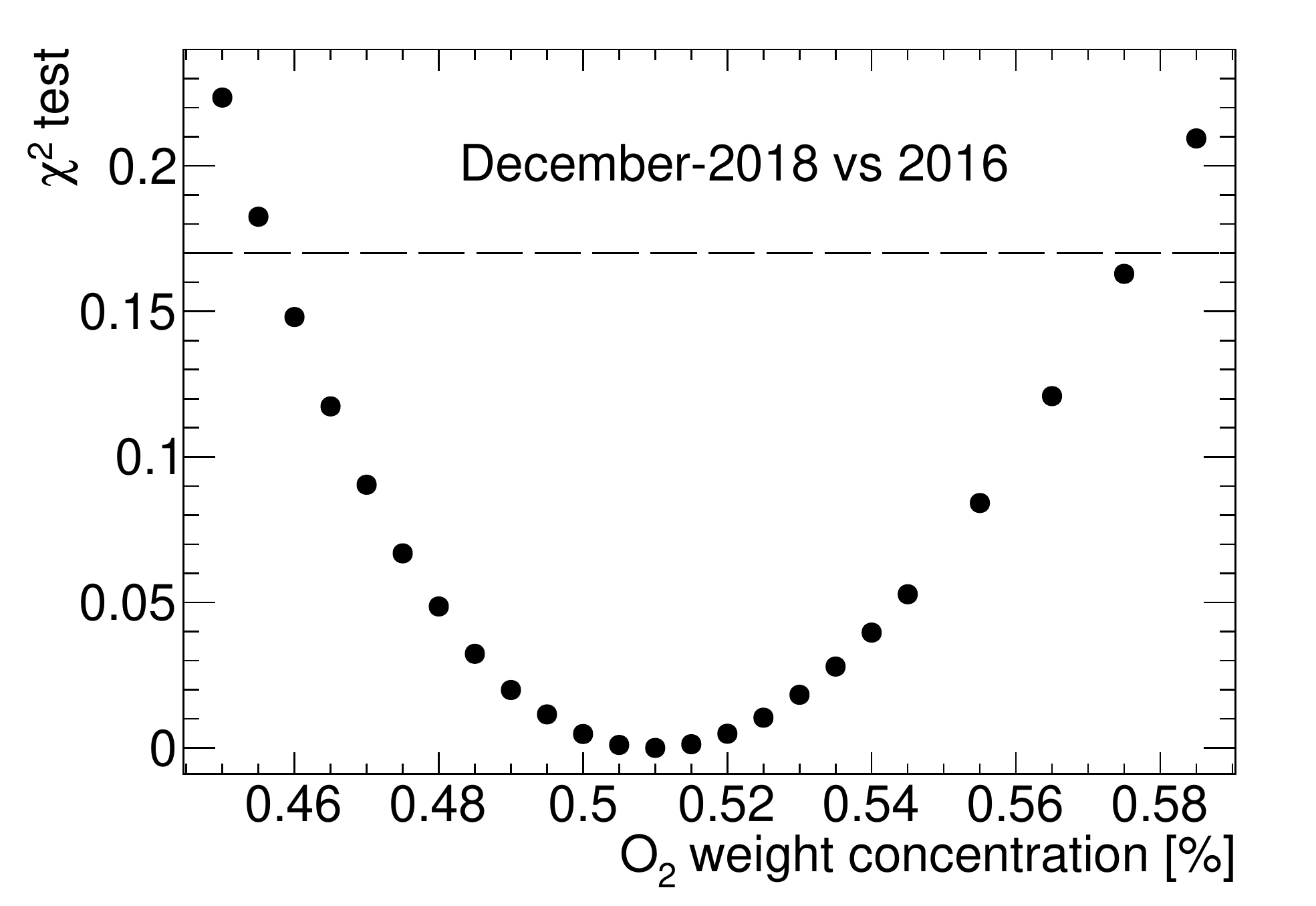}
\centering
\caption{$\chi^2$ of the scaling to 2016 data as a function of the oxygen weight concentration. 
The horizontal line indicates a $\chi^2$ probability of 68$\%$.}
\label{fig:plotscale_tbinw050ns_ndet10}
\end{figure}

The oxygen weight concentration at the minimum $\chi^2$ is \mbox{$c_{O} = 1.61 \pm 0.11 \%$} in March 2018 and \mbox{$c_{O} = 0.51 \pm 0.06 \%$} in December 2018 (see Fig.~\ref{fig:plotscale_tbinw050ns_ndet10}).
The quoted error on $c_{O}$ corresponds to a $\chi^2$ probability of 68$\%$. 
The total systematic uncertainty on $c_{O}$ is obtained by adding in quadrature the relative uncertainty provided by the supplier ($3$\%).
A similar procedure in which the $\chi^2$ of Eq.\ref{eq:chi2} is calculated by comparing directly 2016 and 2018 data points, without any functional fit, leads to consistent results.

\begin{figure}[ht]
\includegraphics[scale=0.3]{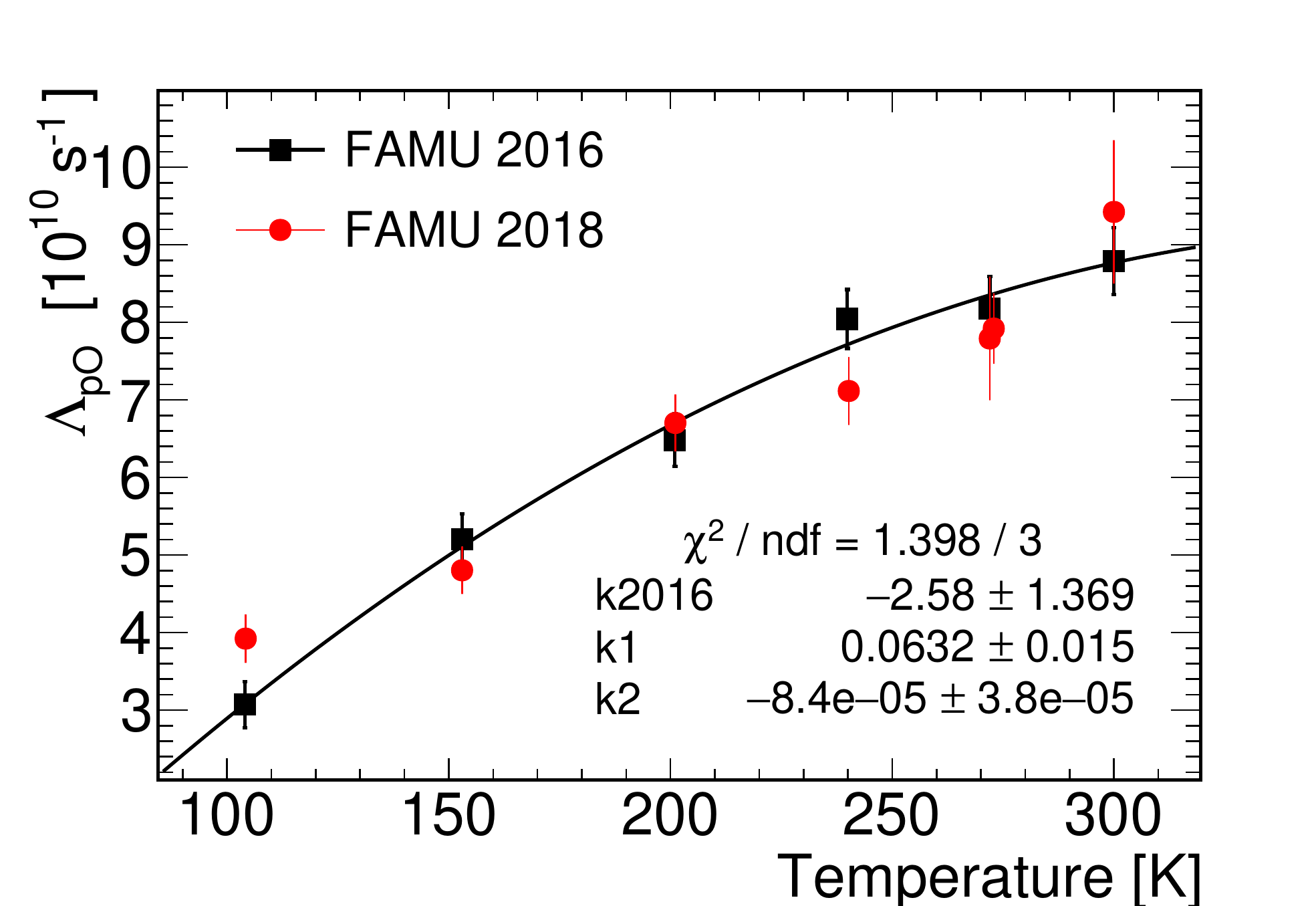}
\centering
\caption{Second degree polynomial fit to the 2016 data \cite{Emiliano} used in the procedure to determine the oxygen weight concentration in 2018.
Normalised 2018 data are superimposed.
The inset shows the reduced $\chi^2$ and the fit parameters, $k1$ and $k2$ being the coefficients of the first and the second degree terms.}
\label{fig:fit2016}
\end{figure}

Figure \ref{fig:fit2016} shows the $2^{nd}$ degree polynomial fit to the 2016 data, together with the 2018 data obtained with $c_{O}$ at the minimum $\chi^2$. 
The December 2018 data are taken at variable temperatures, while March 2018 and 2016 data are taken at a fixed temperature. 
The temperature associated to each 2018 data point is the weighted mean calculated in a range where the temperature decreases linearly with time (more details in Sec.~\ref{sec:data_sample}), the weight being the number of events at a given temperature.
The temperature ranges are chosen in a way that the weighted means correspond to the temperatures of 2016 data points. 

\section{Results} \label{sec:results}

Table \ref{tab:lambda} reports the measurements of \LambdaHO{} performed at constant temperature, with the exception of the value obtained at $70$~K when the temperature was decreasing linearly from $79$~K to $60$~K (more details in Sec.~\ref{sec:data_sample}).

\begin{table}[h]
\centering
\begin{tabular}{ccc}
\hline
\hline
$T$[K] & \LambdaHO[$10^{10} s^{-1}$] & $\chi^2$/ndf \\
\hline
      70     & $^{*}$2.67 $\pm$ 0.40 $\pm$ 0.32 & 1.1 \\ 
      80 $\pm$ 0.5 & 2.96 $\pm$ 0.11 $\pm$ 0.36 & 0.5 \\
     323 $\pm$ 0.5 & 8.88 $\pm$ 0.62 $\pm$ 0.66 & 0.3 \\
     336 $\pm$ 0.5 & 9.37 $\pm$ 0.57 $\pm$ 0.70 & 0.4 \\
\hline
\hline
\end{tabular}
\caption{Measurements of the muon transfer rate to oxygen (\LambdaHO) at different temperatures.
The error on the temperature $T$ indicates the maximum variation of temperature measured with a mK precision during data-taking.
The measurement marked by ($^*$) was performed while the temperature was decreasing from $79$ to $60$~K, 70 K being the event weighted mean.
Additional informations can be found in the Supplementary Material.
The first error on \LambdaHO{} is the sum in quadrature of the statistical uncertainty and the systematic uncertainty associated to background subtraction. 
The second error is the systematic uncertainty associated to the target gas composition. 
The last column reports the reduced $\chi^2$ of the fit to extract \LambdaHO{}.}
\label{tab:lambda}
\end{table}

\begin{figure}[ht]
\includegraphics[scale=0.3]{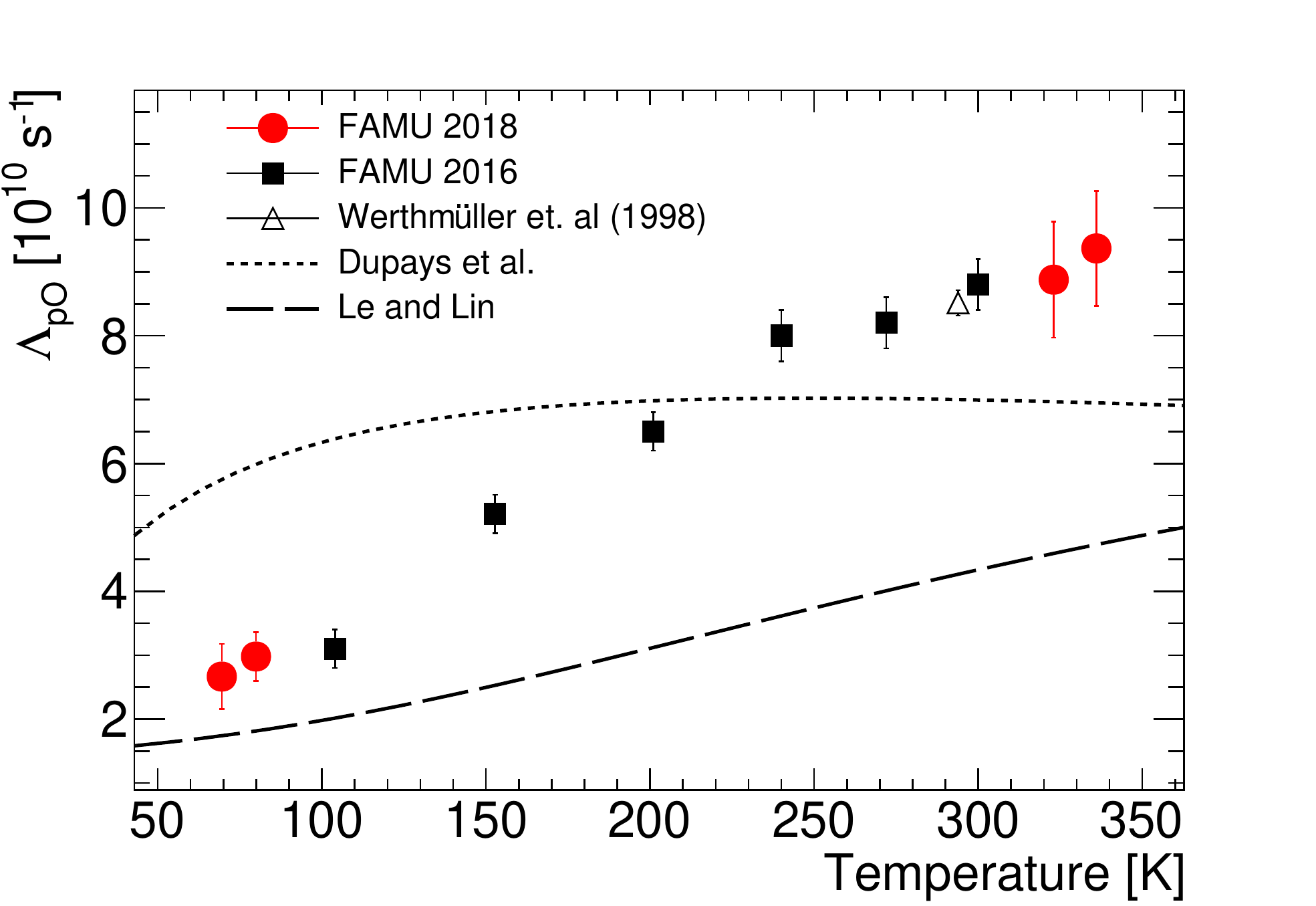}
\centering
\caption{Measurement of \LambdaHO{} as a function of the temperature extracted from 2018 data.
The vertical error bars includes statistical and total systematic effects.
Previous experimental results \cite{Emiliano, Wertmueller_1998} and theoretical predictions \cite{Dupays, LeLin} are superimposed.}
\label{fig:plot2018}
\end{figure}

Figure~\ref{fig:plot2018} shows the results reported in Tab.~\ref{tab:lambda} together with the 2016 measurements \cite{Emiliano}, previous experimental results \cite{Wertmueller_1998} and the predictions of two  theoretical models by Le and Lin \cite{LeLin} and Dupays \cite{Dupays}. 

The 2018 results confirm the rise of \LambdaHO{} with the temperature observed in 2016 and extends the measurements down to $70$~K and up to $336$~K.

The available theoretical predictions do not provide an accurate description of the measurements. However, we observe that a multiplicative factor of 2.0 applied to the model of Ref.\cite{LeLin} leads to a good description of the FAMU data, in particular for the points above 100~K where the $\chi^2/ndf$ is $0.7$.

\newpage

\section{Conclusions}

Data taken with the FAMU detector in 2018 have been analysed to extract the temperature dependence of \LambdaHO{} in the range 70-336~K.
The 2018 measurements have been anchored to the 2016 results in the common temperature ranges by scaling the oxygen concentration of the gas in the target.
The 2018 results confirm the temperature dependence of \LambdaHO{} observed in 2016 and extends the measurements down to 70~K and up to 336~K.
The measurements are in disagreement with the theoretical predictions, as reported already in 2016. 
Since then, new models have been developed but their predictions are not yet public.

\section{Acknowledgements}

The research activity presented in this paper has been carried out in the framework of the FAMU experiment funded by Istituto Nazionale di Fisica Nucleare (INFN).
We thank RAL and the RIKEN-RAL facility for the support and the help in the set-up of the experiment.
We thank the Criotec company for the construction of the FAMU target and for their technical support.
D.B., P.D. and M.S. acknowledge the support from Grant DN 08-17 of the Bulgarian Science Fund.

\bibliography{mybibfile}

\begin{thebibliography}{1}
\expandafter\ifx\csname url\endcsname\relax
  \def\url#1{\texttt{#1}}\fi
\expandafter\ifx\csname urlprefix\endcsname\relax\def\urlprefix{URL }\fi
\expandafter\ifx\csname href\endcsname\relax
  \def\href#1#2{#2} \def\path#1{#1}\fi

\bibitem{Mocchiutti_2018}
{E. Mocchiutti \emph{et al.}}, {{FAMU}: study of the energy dependent transfer
  rate $\Lambda_{\mu p\rightarrow\mu O}$}, Journal of Physics: Conference
  Series 1138 (2018) 012017.
\newblock \href {http://dx.doi.org/10.1088/1742-6596/1138/1/012017}
  {\path{doi:10.1088/1742-6596/1138/1/012017}}.

\bibitem{Emiliano}
{E. Mocchiutti \emph{et al.}}, {First measurement of the temperature dependence
  of muon transfer rate from muonic hydrogen atoms to oxygen}, Phys. Lett. A
  384~(126667).
\newblock \href {http://dx.doi.org/10.1016/j.physleta.2020.126667}
  {\path{doi:10.1016/j.physleta.2020.126667}}.

\bibitem{Cecilia}
{C. Pizzolotto \emph{et al.}}, {The FAMU experiment: muonic hydrogen high
  precision spectroscopy studies}, Eur. Phys. J. A 56 (2020) 185.
\newblock \href {http://dx.doi.org/10.1140/epja/s10050-020-00195-9}
  {\path{doi:10.1140/epja/s10050-020-00195-9}}.

\bibitem{RAL}
{T. Matsuzaki \emph{et al.}}, {The RIKEN-RAL pulsed muon facility}, Nucl.
  Instr. Meth. A 465 (2001) 365.
\newblock \href {http://dx.doi.org/10.1016/S0168-9002(01)00694-5}
  {\path{doi:10.1016/S0168-9002(01)00694-5}}.

\bibitem{Adam2018}
{A. Adamczak \emph{et al.}}, {The FAMU experiment at RIKEN-RAL to study the
  muon transfer rate from hydrogen to other gases}, JINST 13 (2018) P12033.
\newblock \href {http://dx.doi.org/10.1088/1748-0221/13/12/P12033}
  {\path{doi:10.1088/1748-0221/13/12/P12033}}.

\bibitem{Wertmueller_1998}
{Wertm\"uller \emph{et al.}}, {Energy dependence of the charge exchange
  reaction from muonic hydrogen to oxygen}, Hyperfine Interactions 116 (1998)
  1--16.
\newblock \href {http://dx.doi.org/10.1023/A:1012618721239}
  {\path{doi:10.1023/A:1012618721239}}.

\bibitem{LeLin}
A.-T. Le, C.~D. Lin, Muon transfer from muonic hydrogen to atomic oxygen and
  nitrogen, Phys. Rev. A 71 (2005) 022507.
\newblock \href {http://dx.doi.org/10.1103/PhysRevA.71.022507}
  {\path{doi:10.1103/PhysRevA.71.022507}}.

\bibitem{Dupays}
A.~Dupays, B.~Lepetit, J.~A. Beswick, C.~Rizzo, D.~Bakalov, Nonzero
  total-angular-momentum three-body dynamics using hyperspherical elliptic
  coordinates: Application to muon transfer from muonic hydrogen to atomic
  oxygen and neon, Phys. Rev. A 69 (2004) 062501.
\newblock \href {http://dx.doi.org/10.1103/PhysRevA.69.062501}
  {\path{doi:10.1103/PhysRevA.69.062501}}.

\end{thebibliography}

\end{document}